\documentclass[prd,amsfonts,amsmath,twocolumn,superscriptaddress,nofootinbib]{revtex4}

\usepackage{color}
\usepackage{graphicx} 
\usepackage{epstopdf}%
\usepackage{placeins}

\usepackage{dblfloatfix}

\usepackage{subfloat}

\usepackage{mathpazo}      

\newcommand {\be} {\begin{equation}} 
\newcommand {\ba}{\begin{eqnarray}} 
\newcommand {\ee} {\end{equation}} 
\newcommand{\ea} {\end{eqnarray}}

\renewcommand{\epsilon}{\varepsilon}

\begin{document}

\title{Atomic Spectroscopy with Twisted Photons: Separation of M1--E2 Mixed Multipoles}

\author{Andrei Afanasev}

\affiliation{Department of Physics,
The George Washington University, Washington, DC 20052, USA}

\author{Carl E. Carlson}
\affiliation{Department of Physics, The College of William and Mary in Virginia, Williamsburg, VA 23187, USA}

\author{Maria Solyanik$^1$}

\date{\today}

\begin{abstract}
We analyze atomic photoexcitation into the discrete states by twisted photons, or photons carrying extra orbital angular momentum along their direction of propagation. From  the angular momentum and parity considerations, we are able to relate twisted-photon photoexcitation amplitudes to their plane-wave analogues, independently of the details of the atomic wave functions. We analyzed the photo-absorption cross sections of mixed-multipolarity $E2-M1$ transitions in ionized atoms  and found fundamental differences coming from the photon topology.  Our theoretical analysis demonstrates that it is possible to extract the relative transition rates of different multipolar contributions by measuring the photo-excitation rate as a function of the atom's position (or the impact parameter) with respect to the optical vortex center.  The proposed technique for separation of multipoles can be implemented if the target's atom position is resolved with sub-wavelength accuracy, for example, with Paul traps. Numerical examples are presented for Boron-like highly-charged ions (HCI).
\end{abstract}

\maketitle

\section{Introduction}	\label{sec:intro}
Twisted photons, or topological states of light, carrying extra orbital angular momentum (OAM) along their propagation direction, have been one of the trends in optics, photonics and related studies of light-matter interaction for more than 20 years. The seminal paper by Allen \emph{et al.} \cite{allen1992orbital} triggered major development in the field of optical control and manipulation, microscopy, telecommunication, information security, {\it etc.}  In atomic photoexcitation by twisted photons, the terms responsible for vortex behavior often can be conveniently factorized from the conventional plane-wave contribution \cite{afanasev2013off}. Modified atomic selection rules were worked out for Bessel beams (BB)  \cite{afanasev2013off,scholz2014absorption,afanasev2016high}. Later, the formalism was extended to Laguerre-Gaussian (LG) beams \cite{peshkov2017photoexcitation}. The fact that total angular momentum of the twisted photons can be passed to the internal degrees of freedom of an atom was confirmed experimentally for quadrupole transitions with trapped ions \cite{schmiegelow2016transfer,afanasev2017experimental}, and agreed with theoretical predictions. Atomic photo-excitation by vortex beams as a local probe of the beam's topological structure was discussed in Ref. \cite{klimov2012mapping}.  A physics argument for varied strengths for different multipole transitions across the twisted-light wavefront can be found in Ref.\cite{Schmiegelow2012}: While dipole transitions are driven by the electromagnetic field intensity, the quadrupole transitions are caused by the field gradients. For a recent review of interactions between the twisted photons and atoms, see Ref. \cite{Franke-Arnold2017}.

The novel phenomena for the trapped ions interacting with twisted photons have potential applications in quantum computing and quantum storage \cite{northup2014quantum, ruster2017entanglement} due to extra photon-OAM degrees of freedom. Long lifetimes of forbidden states and abundance of nearly degenerate transitions are important benefits for optical clock candidates \cite{safronova2014highly}. Since even a moderate, 3-5\%, increase in lifetimes is important, twisted light can be used as a tool for local high-precision control and tuning of the transition rates. It would allow development of temporal, as well as spatial, measurement techniques based on the knowledge of transition rates with high multipolarity.

Dipole-forbidden transitions are important for measurements of uncertainties in atomic structure and probing physics beyond the standard model \cite{berengut2010enhanced} as well as astrophysics \cite{smitt1976experimental}, and many other related fields, see Ref.~\cite{ludlow2015optical} for review. Transitions forbidden by E1 selection rules recently received a lot of attention in precision spectroscopy \cite{trabert2003m1, beiersdorfer2009spectroscopy, windberger2015identification, safronova2017forbidden}. In this respect, we are particularly interested in studying an interplay of topological properties of the incoming radiation and the atomic system. 


The content of the transitions with mixed multipolarity can be extracted from independent measurements of the transition rates excited by the photons of opposite helicity. Search and tabulation of atomic transitions with mixed multipolarity goes back to 1920's. The technique of using the Zeeman effect to separate different multipolar contributions in such processes is laid out in \cite{mrozowski1944forbidden}. The particular transitions with M1-E2 multipolarity in Bi I, Pb I and II were studied extensively in \cite{kwela1982determination,augustyniak1975zeeman,werbowy2009m1}, where separation of multipolar contributions was analyzed both theoretically and experimentally. The techniques for numerical analysis with extraction of the hyperfine structure for the cases of both integer and half-integer spin were discussed, for instance, in Ref. \cite{wacsowicz2007e2}.

Theoretical description of the photon-atom interaction in a TAM basis is the main focus of this paper. The convenient separation of the total angular momentum (TAM) into orbital and spin parts violates gauge invariance \cite{van1994spin}, which motivates us to work in the TAM basis instead of the conventional linear momentum representation. In the Sec. II we briefly review the quantum mechanical formalism of photo-absorption of the BB and Bessel-Gauss (BG)  light beams by atoms. Sec. III is dedicated to revisiting the QED description of the photon vector potential in TAM basis. In Sec. IV we use this formalism to derive the photo-absorption amplitude  in terms of spherical multipoles for the case of twisted light and discuss the distinct features in the photo-absorption cross section of ions caused by the topology of the incoming beam and OAM transfer to the atomic degrees of freedom. In Sec. V the results are summarized.

\section{Absorption of twisted photons by atoms}\label{sec:2}

In this section we will consider two modes of twisted-light beams: BB and BG. Even though all of them represent optical beam-like fields, they belong to fundamentally different families. BB is an example of non-paraxial mode, structurally stable under propagation. BG is the Helmholtz-type beam which satisfies the paraxial wave equation and is characterized by BB-like behavior close to the beam axis and gaussian-like decay on the periphery.

The fundamental difference between the non-paraxial BB, and paraxial BG modes is that the former one satisfies Maxwell's equations, while the latter ones, strictly speaking, do not. However, one can still apply conventional QED methods to BG modes for the case of not-tightly focused paraxial beams.

Photo-excitation by BB is the convenient place to start due to the elegance of mathematical representation and the property of Bessel functions to form a complete orthonormal basis. These allow us to simply expand the other beam-like solutions in terms of Bessel modes with further application of the developed formalism.

\subsection*{Bessel mode}
As is well-known, e.g. \cite{durnin1987diffraction}, when solving the scalar Helmholtz equation in cylindrical coordinates $(\rho,\phi,z)$, one obtains Bessel mode
\begin{equation}
u^{\text{(BB)}}(\rho, \phi, z; t) = A J_{m_{\gamma}} (\kappa \rho) e^{i(m_\gamma \phi - k_z z)} e^{i\omega t} + c.c.
\end{equation}
where $m_{\gamma}$ is the projection twisted photon's TAM on the direction of propagation; $J_{m}(x)$ is the Bessel function of the first kind, and $k_z$ and $\kappa$ are respectively longitudinal  and transverse components of the photon's wave vector with respect to the propagation direction $z$.  The normalization constant $A=\sqrt{\kappa/2 \pi}$. Following the conventional quantization procedure requires the plane wave expansion of the mode. We use the angular spectrum representation, e.g. \cite{born2013principles}, so that the vector solution can be written as
\begin{flalign}
\hat{A}_{k_z \kappa m_{\gamma} \Lambda} &(\pmb{r}, t) = A \sqrt{\frac{2\pi}{\omega}}\sum_{k}\sum_{\Lambda = -1,1} \int \frac{d^2 k}{(2\pi)^2} a_{\kappa m_{\gamma}} \times &&\nonumber \\
&\times \Big\{ \hat{a}_{k\Lambda} \pmb{e}_{\pmb{k} \Lambda} e^{i(\pmb{k} \cdot \pmb{r} - \omega t)} + \hat{a}_{k\Lambda}^{\dagger} \pmb{e}_{\pmb{k} \Lambda}^{\dagger} e^{-i(\pmb{k} \cdot \pmb{r} - \omega t)} \Big\}&&
\end{flalign}
Here $a_{\kappa m_{\gamma}}$ is the component of the 2D Fourier transform \cite{jentschura2011generation,afanasev2013off}; $\Lambda=\pm 1$ is (spin) helicity of the plane-wave photons forming the Bessel mode and $\pmb{e}_{\pmb{k} \Lambda}$ is the basis state of the twisted-photon polarization, which relates to the plane photon polarization vectors by SO(3) rotation group transformation $\hat{R}_{z}(-\phi_k)\hat{R}_{y}(-\theta_k)\hat{R}_{z}(\phi_k)$ to the linear polarization basis \cite{bliokh2010angular, bliokh2011spin}:
\begin{equation}
\pmb{e}_{\pmb{k} \Lambda} = e^{-i\Lambda \phi_k} \cos^2\frac{\theta_k}{2} \eta_{\Lambda}^{\mu} + e^{i\Lambda \phi_k} \sin^2 \frac{\theta_k}{2} \eta_{-\Lambda}^{\mu} + \frac{\Lambda}{\sqrt{2}} \sin \theta_k \eta_{0}^{\mu}.
\end{equation}
where $\theta_{k}$ is commonly called a pitch angle and $\phi_k$ is the azimuthal angle \cite{afanasev2013off}.  Note that $\Lambda=1(-1)$ corresponds to right- (left-) circular polarization, RCP and LCP. The corresponding local ($e.g.$, at atom's center location) photon flux is
\begin{equation}
\begin{split}
f(\pmb b) = \cos (\theta_k) (|E|^2 + |B|^2)/4 =\;\;\;\;\;\;\;\;\;\;\;\;\;\;\;\;\;\;\;\;\;\;\;\;\;\;\;\;\;\;\\ \cos (\theta_k) \frac{A^2 \omega^2}{2} \Big\{ \cos^4 \frac{\theta_k}{2} J_{m_{\gamma} - \Lambda}^{2}(\kappa b)  +\sin^4 \frac{\theta_k}{2} J_{m_{\gamma} + \Lambda}^{2}(\kappa b) \\+ \frac{\sin \theta_k}{2} J_{m_{\gamma}}^{2}(\kappa b) \Big\}
\end{split}
\label{eq:28_01_2017_1}
\end{equation}
where $\pmb b$ is an {\it impact parameter}, or atom's transverse location with respect to the optical vortex axis. It should be noted that the use of impact-parameter representation for the absorption of twisted light by atoms was demonstrated to be especially practical in Ref.\cite{Kaplan15}.

Proceeding with this approach, the convenient factorization property of the twisted photo-absorption amplitude was obtained, \cite{afanasev2013off,scholz2014absorption,afanasev2016high}:
\begin{flalign}
|M&_{m_f, m_i}^{(\text{BB})} (b)| = \frac{A}{2\pi}|J_{m_f-m_i-m_{\gamma}}(\kappa b) \times &&\nonumber \\
&\times\sum_{m'_f m'_i} d_{m_f, m'_f}^{j_f}(\theta_k) d_{m_i, m'_i}^{j_i}(\theta_k) M_{m'_fm'_i}^{\text{(pw)}}(\theta_k=0)|,&&
\label{29/09/17_2}
\end{flalign}
where $j_{i(f)}$ and $m_{i(f)}$ are TAM of initial (and final) atomic levels and their projections, respectively.
The two terms responsible for the modification of the selection rules are Wigner $d$-functions and Bessel function of the first kind $J_{m_f-m_i-m_{\gamma}}(\kappa b)$, which in the limiting case of a small impact parameter $b \rightarrow 0$ result into the constraint specific for the twisted light: $m_{\gamma} = m_f-m_i$. It implies that at the center of the optical vortex, the TAM projection $m_{\gamma}$ of the incoming photon precisely matches the difference in magnetic quantum numbers of initial and final Zeeman levels, while other transitions are forced to zero by angular momentum conservation. This behavior of twisted-light absorption was demonstrated experimentally with $^{40}$Ca$^+$ ions in a Paul trap \cite{schmiegelow2016transfer,afanasev2017experimental}.

\subsection*{Bessel-Gauss mode}

The BG mode, first discussed in Ref. \cite{sheppard1978gaussian}, is known to be a reliable mathematical representation of real photon laser modes both on the periphery and at the central region. It satisfies the paraxial equation \cite{kiselev2004new} and carries a well-defined TAM:
\begin{equation}
u^{(\text{BG})}(\vec{\rho},t) = A J_n(\kappa \rho) e^{-\rho^2/\text{w}_0^2} e^{im_{\gamma} \phi_{\rho}} e^{-i\omega t} + c.c.
\end{equation}
Making use of the angular spectrum representation, the corresponding photo-excitation amplitude can be expressed as
\begin{flalign}
|M&_{m_f, m_i}^{(\text{BG})} (b)| = e^{-b^2/\omega_0^2} \frac{A}{2\pi}|J_{m_f-m_i-m_{\gamma}}(\kappa b) \times &&\nonumber \\
&\times\sum_{m'_f m'_i} d_{m_f, m'_f}^{j_f}(\theta_k) d_{m_i, m'_i}^{j_i}(\theta_k) M_{m'_fm'_i}^{\text{(pw)}}(\theta_k=0)|,&&
\label{29/09/17_3}
\end{flalign}
For the details of these derivations we refer to the recent paper by Afanasev \emph{et al.} \cite{afanasev2017experimental}.

\section{Expansion in spherical harmonics}\label{sec:3}
As was argued in Ref. \cite{andrews2012angular}, quantum states of non-paraxial beams, such as BB, are not well-defined in the linear momentum basis. Instead the photon's TAM ($j$) basis is used, with minimum $6 \leq 2(2j+1)$ possible states. Vector-potential in terms of spherical multipoles can be defined as
\begin{equation}
\pmb{A}_{jm}^{M}(k, \pmb{r}) = j_j(kr) \pmb{Y}_{jjm}(\Omega);
\label{09/28/2017_1}
\end{equation}
\begin{flalign}
\pmb{A}_{jm}^{E}(k, \pmb{r}) =& \Bigg(\sqrt{\frac{j+1}{2j+1}} j_{j-1}(kr) \pmb{Y}_{j,j-1,m}(\Omega) - &&\nonumber \\
&-\sqrt{\frac{j}{2j+1}} j_{j+1}(kr) \pmb{Y}_{j,j+1,m}(\Omega)\Bigg);
\label{09/28/2017_2}
\end{flalign}
where $M$ and $E$ stand for vector-fields of the magnetic and electric type correspondingly; $j_{m}(x)$ is the spherical Bessel function; and $\pmb{Y}_{j,\ell,m}(\Omega)$ are vector spherical harmonics, e.g. \cite{akhiezer1959quantum}.

In the Coulomb gauge, the non-relativistic quantum mechanical photo-absorption matrix element can be written as
\begin{equation}
S_{fi}^{\text{}} = -i \int dt \; \langle n_f j_f m_f | H_1 | n_i j_i m_i ;\; k \Lambda \rangle;
\end{equation}
where the Hamiltonian operator $H_1$ includes both charge- and spin-dependent parts. For the incoming plane-wave state with well-defined (LCP or RCP) helicity the corresponding matrix element is
\begin{equation}
M_{m_fm_i}^{\text{(pw)}}(\pmb{r}) = \int d\pmb{r}\; \langle n_f j_f m_f | (\hat{p} \cdot \pmb{e}_{\pmb{k} \Lambda}) e^{i\pmb{k} \cdot \pmb{r}} | n_i j_i m_i ;\; k \Lambda \rangle
\label{09/28/2017_3}
\end{equation}
To express the plane-wave photo-absorption amplitude in terms of \eqref{09/28/2017_1} and \eqref{09/28/2017_2} we use the known expansion:
\begin{equation}
\pmb{e}_{\pmb{k} \Lambda} e^{i\pmb{k} \cdot \pmb{r}} =\sqrt{4\pi} \sum_{j=1}^{\infty} \sum_{\ell=j-1}^{j+1} i^{\ell} \sqrt{2\ell+1} j_{\ell}(kr) C_{\ell 0 1\Lambda}^{j\Lambda} \pmb{Y}_{j,\ell, \Lambda}(\Omega)
\end{equation}
After writing out the sum over the projections $\ell$ of TAM of the system, and using the following Clebsch-Gordan coefficients
\begin{equation*}
\begin{gathered}
C_{j-1,0,1,\Lambda}^{j,\Lambda} = \sqrt{\frac{j+1}{2(2j-1)}};\;
C_{j+1,0,1,\Lambda}^{j,\Lambda} = \sqrt{\frac{j}{2(2j+3)}};\\
C_{j,0,1,\Lambda}^{j,\Lambda} = -\frac{\Lambda}{\sqrt{2}};
\end{gathered}
\end{equation*}
one arrives at
\begin{flalign}
\pmb{e}_{\pmb{k} \Lambda} e^{i\pmb{k} \cdot \pmb{r}} =& -\sqrt{4\pi} \sum_{j=1}^{\infty} \sqrt{\frac{(2j+1)}{2}} i^j  \times &&\nonumber \\
& \times \Big\{ i \pmb{A}_{jm}^{E}(k, \pmb{r}) + \Lambda \pmb{A}_{jm}^{M}(k, \pmb{r})\Big\};
\end{flalign}
This expansion now can be used in \eqref{09/28/2017_3}
\begin{equation}
M_{m_fm_i}^{\text{(pw)}}(0) = -\sqrt{4\pi} \sum_{j=1}^{\infty} i^{j+\mu} \sqrt{\tfrac{(2j+1)}{(2 j_f+1)}} \Lambda^{\mu+1} C_{j_i,m_i,j,\Lambda}^{j_f, m_f}\;M_{j\mu}
\label{10/5/17_1}
\end{equation}
such that $\mu=1$ stands for electric, and $\mu=0$ for magnetic multipolarity. Here $M_{j\mu}$ stands for the spherical amplitudes of multipolarity $\mu$ and order $j$. Substituting this transition amplitude into the factorization formulas \eqref{29/09/17_2}, and \eqref{29/09/17_3}, we express the photo-absorption amplitude in terms of electric and magnetic multipoles.

\section{Photo-excitations in highly charged ions}\label{sec:4}

According to recent theoretical and experimental studies, Boron-like and Sn-like HCI's are among the best candidates for the next generation of atomic clocks. At the same time, spectral lines of HCI are commonly characterized by mixed multipolarity. In this section we use HCIs to demonstrate distinctive features arising from the OAM transfer from the photon and analyze a possibility of separation of multipoles with OAM light.

The photoexcitation rate $\Gamma$ and cross section $\sigma$ can be obtained from the above formulae for the transition matrix elements $c.f.$ \cite{afanasev2013off,afanasev2017circular} as
\begin{equation}
\begin{gathered}
\Gamma^{\text{(tw)}}_{M_{j \mu}}(b) = 2\pi \delta (E_f-E_i-\omega) \sum_{m_f m_i} |M^{(\text{tw})}_{m_f m_i}|^2 ,\\
\sigma^{\text{(tw)}}_{M_{j \mu}}(b)=\Gamma^{\text{(tw)}}_{M_{j \mu}}(b)/f(\pmb b).
\end{gathered}
\end{equation}
where  $f(\pmb b)$ is the local (b-dependent) flux, as in Eq. \eqref{eq:28_01_2017_1} for Bessel mode. 

Analyzing eqn. \eqref{29/09/17_2} or \eqref{29/09/17_3}, one can see that the cross section in the most general case is
\begin{flalign}
&\sigma^{\text{(tw)}}_{M_{j \mu}}(\pmb{b}) = 2\pi \delta (E_f-E_i-\omega) \frac{A^2}{f(\pmb b)} \times &&\nonumber \\ 
&\times \sum_{m_f m_i} \sum_{ m'_f m'_i j} |J_{m_f - m_{\gamma}-m_i}(\kappa b) d_{m_f m'_f}^{j_f} d_{m_i m'_i}^{j_i} \mathcal{C}_{jj_fj_i} M_{j\mu}|^2&&
\label{12/25/2017_1}
\end{flalign}
where $\mathcal{C}_{jj_fj_i}$ is the coefficient in \eqref{10/5/17_1}:
\begin{equation}
\mathcal{C}_{jj_fj_i} = -\sqrt{4\pi} \; i^{j+\mu} \sqrt{\tfrac{(2j+1)}{(2 j_f+1)}} \Lambda^{\mu+1} C_{j_i,m_i,j,\Lambda}^{j_f, m_f}
\end{equation}

We will start from the low-level transitions with definite multipolarity and initial atomic TAM $j_i = 0$.  One can verify that a distinctive feature of the twisted light photo-absorption is that it relaxes the plane wave selection rules: instead of having only one allowed amplitude for $\Delta m=m_f-m_i\Lambda$, we get $2\Delta j+1$ amplitudes possible per process, where $\delta_j=j_f-j_i$. The photo-absorption cross section for BB \eqref{12/25/2017_1} takes the form ($c.f.$ Refs.\cite{afanasev2016high,scholz2014absorption} that neglected electron's spin):

\begin{flalign}
\sigma^{\text{(tw)}}_{M_{j \mu}}(\pmb{b}) &= 2\pi \delta (E_f-E_i-\omega) \frac{A^2}{f(\pmb{b})} \times &&\nonumber \\ 
&\times \sum_{m_f}|C_{00j_f \Lambda}^{j_fm_f}J_{m_f - m_{\gamma}}(\kappa b) d_{m_f \Lambda}^{j_f}(\theta_k) M_{j_f \mu}|^2&&
\end{flalign}
which, for the case of $j_f=1$, can be shown to be proportional to incoming photon flux $\sigma_{m_f}(\pmb{b}) \propto |M_{10}^{\text{(pw)}}|^2$. Clebsch-Gordan coefficients are coming from coupling of the EM-field topological charge to the internal degrees of freedom of the HCI. The multipolarity is determined by the TAM exchange $\Delta j$ and the parity of the final state. This leads to the conclusion that E1 and M1 transition rates with the twisted photons are factorizable and proportional to the intensity of the incoming radiation. For the case of multipoles of higher $\Delta j$, the characteristic factorization is also present, but the excitation rates acquire extra terms, proportional to $J_{m_{\gamma} \pm c}^2 (\kappa b)$, where $1<c \leq \Delta j$.

\begin{figure}[h]
\centering
\includegraphics[scale=.65]{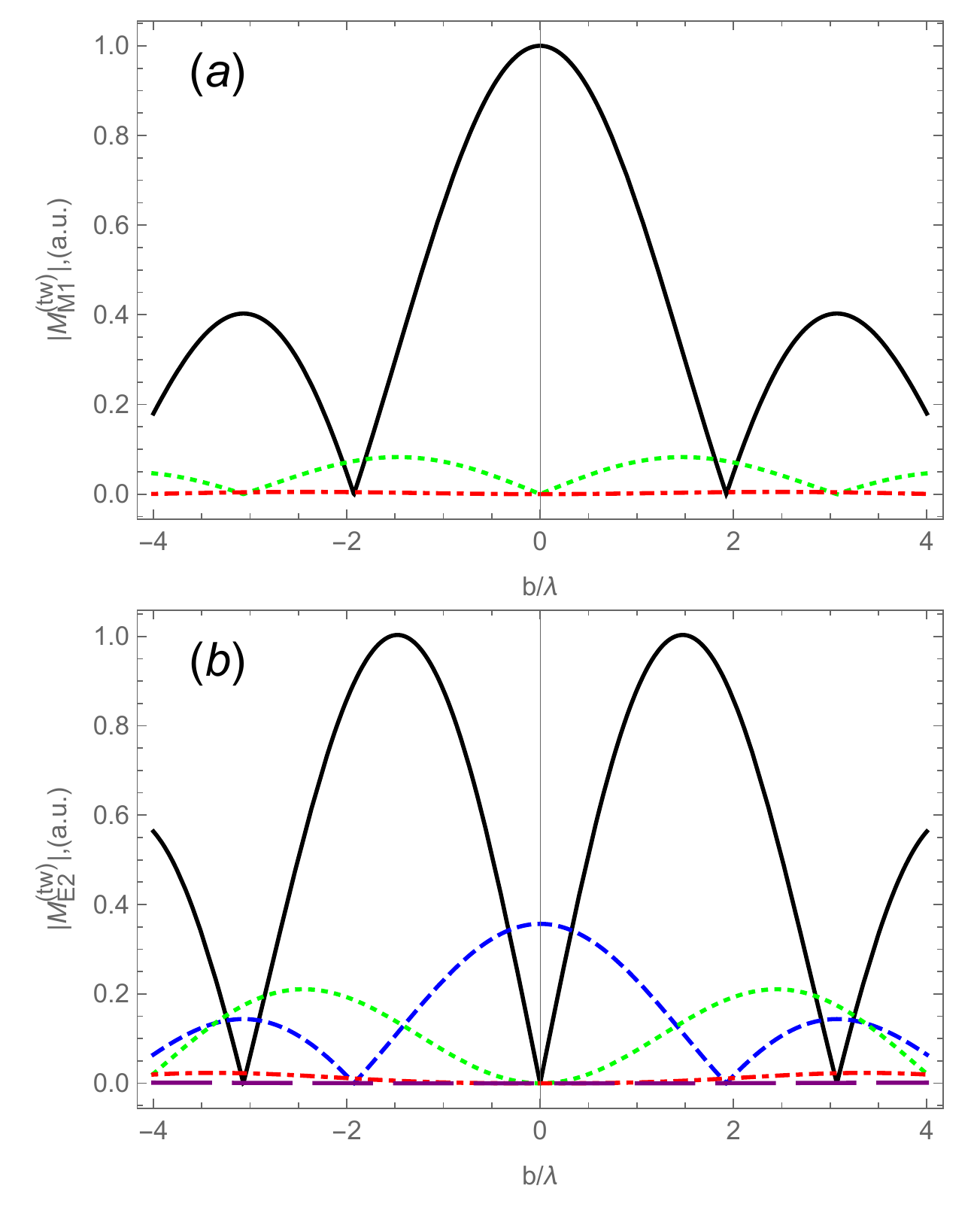}
\caption{Dependence of photo-absorption amplitudes of M1($\lambda$=351 nm) for $m_{\gamma} = 1$ (a) and E2($\lambda$=426 nm) for $m_{\gamma} = 2$ (b) transitions in Pr$^{9+}$ HCI of OAM photons with Bessel profile for $\Delta m=2$ - dashed-blue, $\Delta m=1$ - black-solid, $\Delta m=0$ - dotted-green, $\Delta m=-1$ - dot-dashed-red, $\Delta m=-2$ - long-dashed-purple. $\Lambda=1$ (RCP) in both plots.}
\label{Pic1}
\end{figure}

\begin{figure}[h!]
\centering
\includegraphics[scale=.65]{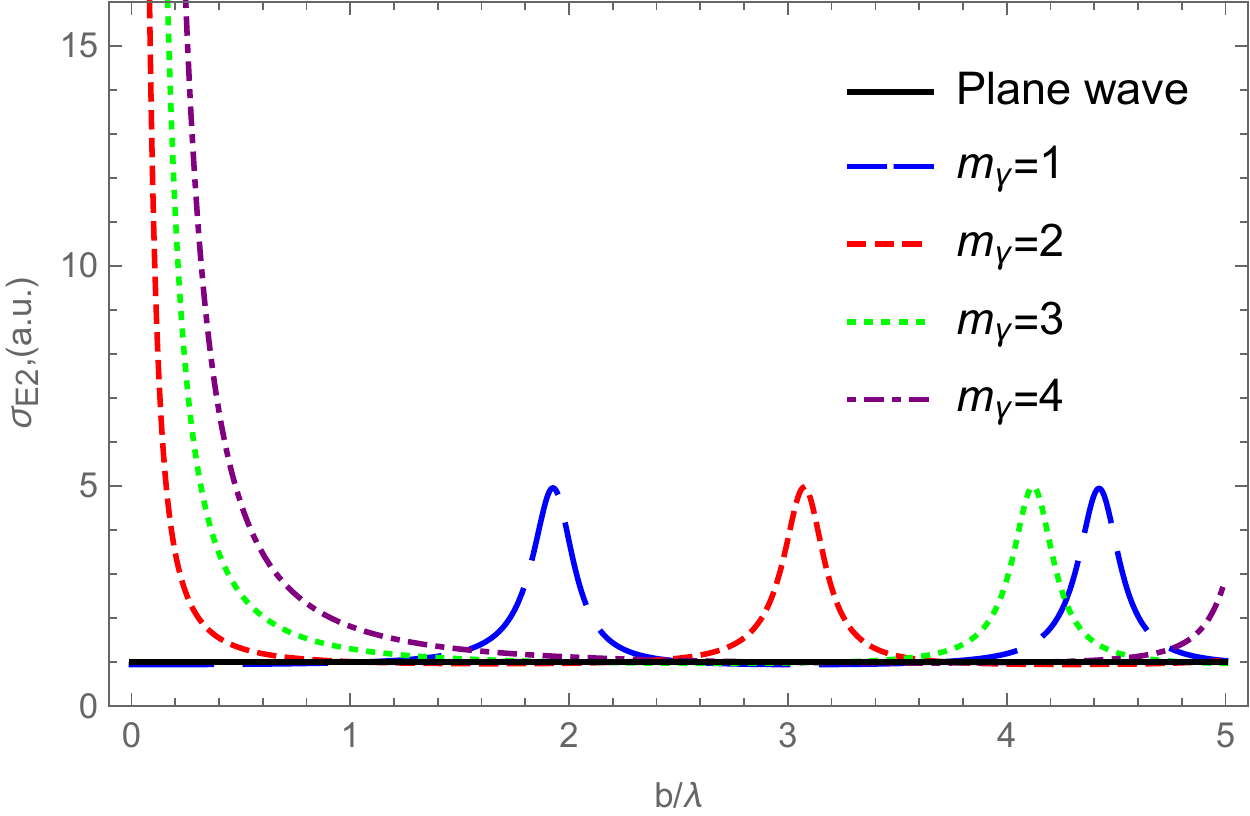}
\caption{Photo-absorption cross section for E2 in Pr$^{9+}$ HCI by twisted photons with Bessel profile for pitch angle $\theta_k=0.2$, $\Lambda=1$ (RCP)  and different TAM projections $m_{\gamma}$. The black solid line is the plane-wave cross section normalized to one, see \cite{safronova2014highly} for actual lifetimes.}
\label{Pic2}
\end{figure}

As an example, let us consider transitions from the ground level  in Sn-like Pr$^{9+}$: 1) 351 nm M1 transition $^3P_0 \rightarrow \;^3P_1$; 2) 426 nm E2 transition $^3P_0 \rightarrow \;^3F_2$. Corresponding amplitudes for the twisted photons for both M1 and E2 transitions are shown in Fig.\ref{Pic1} for BB profile. The use of BG profile suppresses the amplitudes at beam periphery depending on the choice of waist parameter. Squaring the transition amplitudes, after summation the rate of $M1$ transition $\Gamma^{(\text{tw})}_{M1}=f \cdot \sigma^{(\text{tw})}_{M1}(\pmb b)$ appears to be proportional to the electromagnetic flux at the given impact parameter $b$, making the cross section independent on atom's position. However, the electric quadrupole cross section shows the characteristic periodic pattern of peaks, dependent on the impact parameter of the system and the pitch angle $\theta_k$. Formally, the cross section for E2 transition becomes singular in the optical vortex center, a phenomenon that can be related to "excitation in the dark" demonstrated experimentally in Refs.\cite{schmiegelow2016transfer}, see also Ref.\cite{afanasev2016high} for theory discussion. The photoexcitation cross section for E2-transition as a function of impact parameter $b$ is illustrated in Fig.\ref{Pic2}. 

\FloatBarrier

Next, we considered transitions with mixed multipolarity, such as $^2P_{1/2} \rightarrow \;^2D_{3/2}$ ($142$nm) in Boron-like atoms, where the overall local transition amplitude comes from both E2 and M1-type contributions, while E1 transitions are parity-forbidden. It was calculated by Rynkun et al. \cite{rynkun2012energies} that the magnetic dipole contribution is slightly larger than that of the electric quadrupole in these transitions, $M1/E2 \approx 1.1$. This makes it especially convenient for studying the effects coming from photon topology in mixed-multipolarity states. For the plane-wave case we get two allowed transitions, where relative normalization of the multipoles follows from Eq.(\ref{12/25/2017_1}):
\begin{equation}
\begin{gathered}
M_{3/2,1/2}^{\text{(pw)}}(0) = i \sqrt{\pi} (E2-\sqrt{3} M1)\\
M_{1/2,-1/2}^{\text{(pw)}}(0) = -i \sqrt{\pi} (\sqrt{3}E2+M1).
\end{gathered}  \label{12/25/2017_2}
\end{equation}
For the twisted photons, one can check that the plane wave amplitudes do not factorize out in this case. However, the whole expression for the OAM cross section remains free of the interference terms.
\begin{figure}[h!]
\centering
\includegraphics[scale=.55]{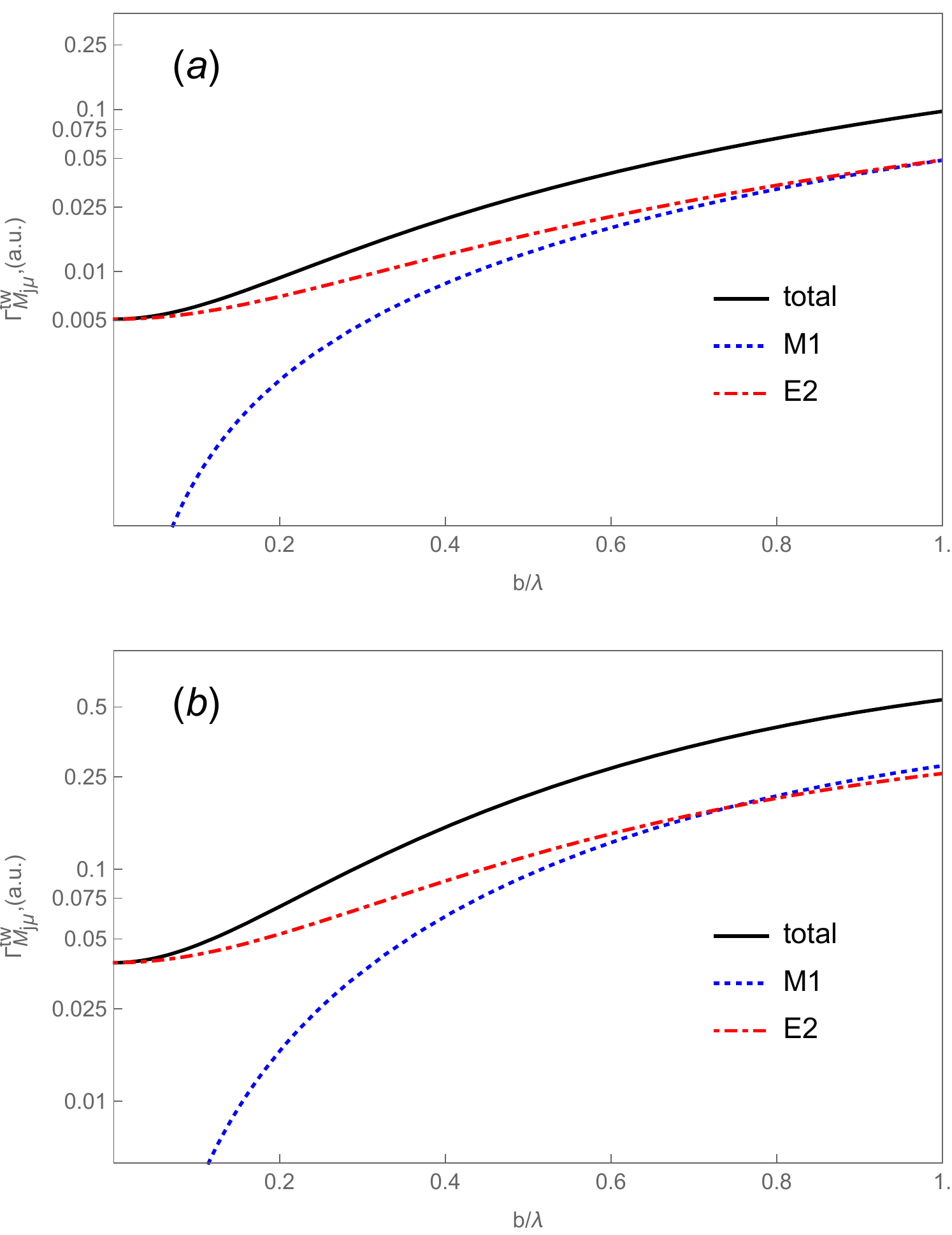}
\caption{Log-plots of photo-absorption rates in Boron-like HCI for pitch angles $\theta_k=0.1$ and 0.2. The transitions are excited by twisted photons with Bessel profile, $m_{\gamma}=2$, and right-handed helicity ($\Lambda=1$).}
\label{Pic3}
\end{figure}
This allows us to use the following equation for the local photo-absorption cross section:
\begin{flalign}
&\sigma^{\text{(tw)}}_{M_{j \mu}}(\pmb{b}) = 2\pi \delta (E_f-E_i-\omega) \frac{A^2}{f(\pmb b)} \times &&\nonumber \\ 
&\times \sum_{m_f m_i} \sum_{ m'_f m'_i j} |J_{m_f - m_{\gamma}-m_i}(\kappa b) d_{m_f m'_f}^{3/2} d_{m_i m'_i}^{1/2} \mathcal{C}_{j3/2 \;1/2}|^2 |M_{j\mu}|^2&&
\label{12/25/2017_3}
\end{flalign}
where $M_{j\mu}$ are the multipoles from Eq. \eqref{12/25/2017_2}. 

Treating the $\theta_k$ as a small parameter $\theta_k \rightarrow 0$, one can expand the expression \eqref{12/25/2017_1} with local flux $f(\pmb{b})$, eqn.\eqref{eq:28_01_2017_1} and get the leading terms of the expansion as
\begin{gather}
\sigma^{\text{(tw)}}_{m_{\gamma} = 1} \rightarrow 4\pi(E2^2+M1^2)+O(\theta_k^2);\\
\sigma^{\text{(tw)}}_{m_{\gamma} = 2} \rightarrow \Big(4\pi(E2^2 + M1^2)+\frac{4 E2^2}{(b/\lambda)^2 \pi}\Big) +O(\theta_k^2);\\
\sigma^{\text{(tw)}}_{m_{\gamma} = 3} \rightarrow  \Big(4\pi (E2^2+M1^2)+ \frac{16 E2^2}{(b/\lambda)^2 \pi}\Big)+ O(\theta_k^2);\\
\sigma^{\text{(pw)}}=4\pi(E2^2+M1^2).
\end{gather}
where the leading multipole contribution $4\pi (E2^2+M1^2)$ corresponds to the plane wave cross section $\sigma^{(\text{pw})}$. Due to the factorization property of BG amplitudes,
the derived expression for cross sections apply both for BB and BG modes.

In Fig.~\ref{Pic3} the photoexcitation rates for these transitions is plotted as a function of the impact parameter $b$, where the individual contributions from E2- and M1-transitions, and their total are shown for BB profile. Comparing results for two different values of the pitch angle $\theta_k$ we find that the rate is smaller for smaller $\theta_k$, which can be understood as the effect of Wigner functions in Eqs.(16,20). The use of BG profile suppresses the rate on beam periphery, but does not affect relative contributions of M1 and E2 multipoles. One can see the strong domination of the electric quadrupole over the magnetic dipole in the center of the beam. The effect becomes noticeable for the distances $b\leq\lambda/3$, which for the considered case of $\lambda = 142$nm is about $50$nm. It imposes position-resolution requirements on the possible experimental observation of the predicted effect.

\section{Summary and Outlook}\label{sec:5}

In this paper we presented a theoretical description of the multipolar structure of the twisted light based on the fundamental representation of the photon in its TAM basis. The atomic photo-excitation amplitude is obtained in the form of the multipole expansion.

We analyzed the photo-absorption cross sections of mixed $E2-M1$ transitions in ionized atoms interacting with OAM-photons, which revealed fundamental differences coming from the photon topology. Two distinct and novel features of the twisted-photon photoexcitation are observed: (a) The magnetic levels population is strongly affected by the topological charge of the photons and (b) The relative contributions of the M1-, E2-amplitudes into the mixed-multipole transitions depend on the atom's location with respect to the optical vortex axis. According to our theoretical analysis, it is possible to extract the relative transition rates of different multipolar contributions by measuring the photo-excitation rate as a function of the atom's position (or the impact parameter) with respect to the optical vortex center. In this case, only the $E2$ transition survives at the vortex center for the incoming photons carrying two units of angular momentum along the propagation direction. On the other hand, the rates at the beam's periphery are driven by the same relative contribution of multipoles as in the plane-wave case. The proposed method of multipole separation with twisted light requires high position resolution of the target atom's position that can be provided, for example, by Paul traps, as in the recent experiments with $^{40}$Ca$^+$ ions \cite{schmiegelow2016transfer,afanasev2017experimental}. 

In addition, experimental implementation of the proposed technique for HCI would require a source of the twisted light in UV range. Presently, generation of the twisted light up to XUV range (with 99~eV photons) was demonstrated at the synchrotron light source BESSY II \cite{BESSY99ev}, and it is feasible with new-generation light sources.

\section*{Acknowlegements}

Work of A.A. and M.S. was supported in part by Gus Weiss Endowment of George Washington University. CEC thanks the National Science Foundation for support under Grant PHY-1516509 and thanks the Johannes Gutenberg-University, Mainz, for hospitality as this work was being completed. The authors would like to thank Ferdinand Schmidt-Kaler, Christian Schmiegelow and Valery Serbo for useful discussions.

\bibliography{Master_b} 

\begin{thebibliography}{39}
\expandafter\ifx\csname natexlab\endcsname\relax\def\natexlab#1{#1}\fi
\expandafter\ifx\csname bibnamefont\endcsname\relax
  \def\bibnamefont#1{#1}\fi
\expandafter\ifx\csname bibfnamefont\endcsname\relax
  \def\bibfnamefont#1{#1}\fi
\expandafter\ifx\csname citenamefont\endcsname\relax
  \def\citenamefont#1{#1}\fi
\expandafter\ifx\csname url\endcsname\relax
  \def\url#1{\texttt{#1}}\fi
\expandafter\ifx\csname urlprefix\endcsname\relax\def\urlprefix{URL }\fi
\providecommand{\bibinfo}[2]{#2}
\providecommand{\eprint}[2][]{\url{#2}}

\bibitem[{\citenamefont{Allen et~al.}(1992)\citenamefont{Allen, Beijersbergen,
  Spreeuw, and Woerdman}}]{allen1992orbital}
\bibinfo{author}{\bibfnamefont{L.}~\bibnamefont{Allen}},
  \bibinfo{author}{\bibfnamefont{M.~W.} \bibnamefont{Beijersbergen}},
  \bibinfo{author}{\bibfnamefont{R.}~\bibnamefont{Spreeuw}}, \bibnamefont{and}
  \bibinfo{author}{\bibfnamefont{J.}~\bibnamefont{Woerdman}},
  \bibinfo{journal}{Physical Review A} \textbf{\bibinfo{volume}{45}},
  \bibinfo{pages}{8185} (\bibinfo{year}{1992}).

\bibitem[{\citenamefont{Afanasev et~al.}(2013)\citenamefont{Afanasev, Carlson,
  and Mukherjee}}]{afanasev2013off}
\bibinfo{author}{\bibfnamefont{A.}~\bibnamefont{Afanasev}},
  \bibinfo{author}{\bibfnamefont{C.~E.} \bibnamefont{Carlson}},
  \bibnamefont{and}
  \bibinfo{author}{\bibfnamefont{A.}~\bibnamefont{Mukherjee}},
  \bibinfo{journal}{Physical Review A} \textbf{\bibinfo{volume}{88}},
  \bibinfo{pages}{033841} (\bibinfo{year}{2013}).

\bibitem[{\citenamefont{Scholz-Marggraf
  et~al.}(2014)\citenamefont{Scholz-Marggraf, Fritzsche, Serbo, Afanasev, and
  Surzhykov}}]{scholz2014absorption}
\bibinfo{author}{\bibfnamefont{H.}~\bibnamefont{Scholz-Marggraf}},
  \bibinfo{author}{\bibfnamefont{S.}~\bibnamefont{Fritzsche}},
  \bibinfo{author}{\bibfnamefont{V.}~\bibnamefont{Serbo}},
  \bibinfo{author}{\bibfnamefont{A.}~\bibnamefont{Afanasev}}, \bibnamefont{and}
  \bibinfo{author}{\bibfnamefont{A.}~\bibnamefont{Surzhykov}},
  \bibinfo{journal}{Physical Review A} \textbf{\bibinfo{volume}{90}},
  \bibinfo{pages}{013425} (\bibinfo{year}{2014}).

\bibitem[{\citenamefont{Afanasev et~al.}(2016)\citenamefont{Afanasev, Carlson,
  and Mukherjee}}]{afanasev2016high}
\bibinfo{author}{\bibfnamefont{A.}~\bibnamefont{Afanasev}},
  \bibinfo{author}{\bibfnamefont{C.~E.} \bibnamefont{Carlson}},
  \bibnamefont{and}
  \bibinfo{author}{\bibfnamefont{A.}~\bibnamefont{Mukherjee}},
  \bibinfo{journal}{Journal of Optics} \textbf{\bibinfo{volume}{18}},
  \bibinfo{pages}{074013} (\bibinfo{year}{2016}).

\bibitem[{\citenamefont{Peshkov et~al.}(2017)\citenamefont{Peshkov, Seipt,
  Surzhykov, and Fritzsche}}]{peshkov2017photoexcitation}
\bibinfo{author}{\bibfnamefont{A.}~\bibnamefont{Peshkov}},
  \bibinfo{author}{\bibfnamefont{D.}~\bibnamefont{Seipt}},
  \bibinfo{author}{\bibfnamefont{A.}~\bibnamefont{Surzhykov}},
  \bibnamefont{and}
  \bibinfo{author}{\bibfnamefont{S.}~\bibnamefont{Fritzsche}},
  \bibinfo{journal}{Physical Review A} \textbf{\bibinfo{volume}{96}},
  \bibinfo{pages}{023407} (\bibinfo{year}{2017}).

\bibitem[{\citenamefont{Schmiegelow et~al.}(2016)\citenamefont{Schmiegelow,
  Schulz, Kaufmann, Ruster, Poschinger, and
  Schmidt-Kaler}}]{schmiegelow2016transfer}
\bibinfo{author}{\bibfnamefont{C.~T.} \bibnamefont{Schmiegelow}},
  \bibinfo{author}{\bibfnamefont{J.}~\bibnamefont{Schulz}},
  \bibinfo{author}{\bibfnamefont{H.}~\bibnamefont{Kaufmann}},
  \bibinfo{author}{\bibfnamefont{T.}~\bibnamefont{Ruster}},
  \bibinfo{author}{\bibfnamefont{U.~G.} \bibnamefont{Poschinger}},
  \bibnamefont{and}
  \bibinfo{author}{\bibfnamefont{F.}~\bibnamefont{Schmidt-Kaler}},
  \bibinfo{journal}{Nature Communications} \textbf{\bibinfo{volume}{7}}
  (\bibinfo{year}{2016}).

\bibitem[{\citenamefont{Afanasev et~al.}(2018)\citenamefont{Afanasev, Carlson,
  Schmiegelow, Schulz, Schmidt-Kaler, and Solyanik}}]{afanasev2017experimental}
\bibinfo{author}{\bibfnamefont{A.}~\bibnamefont{Afanasev}},
  \bibinfo{author}{\bibfnamefont{C.~E.} \bibnamefont{Carlson}},
  \bibinfo{author}{\bibfnamefont{C.~T.} \bibnamefont{Schmiegelow}},
  \bibinfo{author}{\bibfnamefont{J.}~\bibnamefont{Schulz}},
  \bibinfo{author}{\bibfnamefont{F.}~\bibnamefont{Schmidt-Kaler}},
  \bibnamefont{and} \bibinfo{author}{\bibfnamefont{M.}~\bibnamefont{Solyanik}},
  \bibinfo{journal}{New Journal of Physics}  (\bibinfo{year}{2018}),
  \urlprefix\url{http://iopscience.iop.org/article/10.1088/1367-2630/aaa63d}.

\bibitem[{\citenamefont{Klimov et~al.}(2012)\citenamefont{Klimov, Bloch,
  Ducloy, and Leite}}]{klimov2012mapping}
\bibinfo{author}{\bibfnamefont{V.~V.} \bibnamefont{Klimov}},
  \bibinfo{author}{\bibfnamefont{D.}~\bibnamefont{Bloch}},
  \bibinfo{author}{\bibfnamefont{M.}~\bibnamefont{Ducloy}}, \bibnamefont{and}
  \bibinfo{author}{\bibfnamefont{J.~R.} \bibnamefont{Leite}},
  \bibinfo{journal}{Physical Review A} \textbf{\bibinfo{volume}{85}},
  \bibinfo{pages}{053834} (\bibinfo{year}{2012}).

\bibitem[{\citenamefont{Schmiegelow and Schmidt-Kaler}(2012)}]{Schmiegelow2012}
\bibinfo{author}{\bibfnamefont{C.}~\bibnamefont{Schmiegelow}} \bibnamefont{and}
  \bibinfo{author}{\bibfnamefont{F.}~\bibnamefont{Schmidt-Kaler}},
  \bibinfo{journal}{European Physics Journal D} \textbf{\bibinfo{volume}{66}},
  \bibinfo{pages}{157} (\bibinfo{year}{2012}),
  \urlprefix\url{http://dx.doi.org/10.1140/epjd/e2012-20730-4}.

\bibitem[{\citenamefont{Franke-Arnold}(2017)}]{Franke-Arnold2017}
\bibinfo{author}{\bibfnamefont{S.}~\bibnamefont{Franke-Arnold}},
  \bibinfo{journal}{Philosophical Transactions of the Royal Society of London
  A: Mathematical, Physical and Engineering Sciences}
  \textbf{\bibinfo{volume}{375}} (\bibinfo{year}{2017}), ISSN
  \bibinfo{issn}{1364-503X},
  \urlprefix\url{http://rsta.royalsocietypublishing.org/content/375/2087/20150435}.

\bibitem[{\citenamefont{Northup and Blatt}(2014)}]{northup2014quantum}
\bibinfo{author}{\bibfnamefont{T.}~\bibnamefont{Northup}} \bibnamefont{and}
  \bibinfo{author}{\bibfnamefont{R.}~\bibnamefont{Blatt}},
  \bibinfo{journal}{Nature Photonics} \textbf{\bibinfo{volume}{8}},
  \bibinfo{pages}{356} (\bibinfo{year}{2014}).

\bibitem[{\citenamefont{Ruster et~al.}(2017)\citenamefont{Ruster, Kaufmann,
  Luda, Kaushal, Schmiegelow, Schmidt-Kaler, and
  Poschinger}}]{ruster2017entanglement}
\bibinfo{author}{\bibfnamefont{T.}~\bibnamefont{Ruster}},
  \bibinfo{author}{\bibfnamefont{H.}~\bibnamefont{Kaufmann}},
  \bibinfo{author}{\bibfnamefont{M.}~\bibnamefont{Luda}},
  \bibinfo{author}{\bibfnamefont{V.}~\bibnamefont{Kaushal}},
  \bibinfo{author}{\bibfnamefont{C.}~\bibnamefont{Schmiegelow}},
  \bibinfo{author}{\bibfnamefont{F.}~\bibnamefont{Schmidt-Kaler}},
  \bibnamefont{and}
  \bibinfo{author}{\bibfnamefont{U.}~\bibnamefont{Poschinger}},
  \bibinfo{journal}{arXiv preprint arXiv:1704.01793}  (\bibinfo{year}{2017}).

\bibitem[{\citenamefont{Safronova et~al.}(2014)\citenamefont{Safronova, Dzuba,
  Flambaum, Safronova, Porsev, and Kozlov}}]{safronova2014highly}
\bibinfo{author}{\bibfnamefont{M.}~\bibnamefont{Safronova}},
  \bibinfo{author}{\bibfnamefont{V.}~\bibnamefont{Dzuba}},
  \bibinfo{author}{\bibfnamefont{V.}~\bibnamefont{Flambaum}},
  \bibinfo{author}{\bibfnamefont{U.}~\bibnamefont{Safronova}},
  \bibinfo{author}{\bibfnamefont{S.}~\bibnamefont{Porsev}}, \bibnamefont{and}
  \bibinfo{author}{\bibfnamefont{M.}~\bibnamefont{Kozlov}},
  \bibinfo{journal}{Physical Review Letters} \textbf{\bibinfo{volume}{113}},
  \bibinfo{pages}{030801} (\bibinfo{year}{2014}).

\bibitem[{\citenamefont{Berengut et~al.}(2010)\citenamefont{Berengut, Dzuba,
  and Flambaum}}]{berengut2010enhanced}
\bibinfo{author}{\bibfnamefont{J.}~\bibnamefont{Berengut}},
  \bibinfo{author}{\bibfnamefont{V.}~\bibnamefont{Dzuba}}, \bibnamefont{and}
  \bibinfo{author}{\bibfnamefont{V.}~\bibnamefont{Flambaum}},
  \bibinfo{journal}{Physical Review Letters} \textbf{\bibinfo{volume}{105}},
  \bibinfo{pages}{120801} (\bibinfo{year}{2010}).

\bibitem[{\citenamefont{Smitt et~al.}(1976)\citenamefont{Smitt, Svensson, and
  Outred}}]{smitt1976experimental}
\bibinfo{author}{\bibfnamefont{R.}~\bibnamefont{Smitt}},
  \bibinfo{author}{\bibfnamefont{L.~{\AA}.} \bibnamefont{Svensson}},
  \bibnamefont{and} \bibinfo{author}{\bibfnamefont{M.}~\bibnamefont{Outred}},
  \bibinfo{journal}{Physica Scripta} \textbf{\bibinfo{volume}{13}},
  \bibinfo{pages}{293} (\bibinfo{year}{1976}).

\bibitem[{\citenamefont{Ludlow et~al.}(2015)\citenamefont{Ludlow, Boyd, Ye,
  Peik, and Schmidt}}]{ludlow2015optical}
\bibinfo{author}{\bibfnamefont{A.~D.} \bibnamefont{Ludlow}},
  \bibinfo{author}{\bibfnamefont{M.~M.} \bibnamefont{Boyd}},
  \bibinfo{author}{\bibfnamefont{J.}~\bibnamefont{Ye}},
  \bibinfo{author}{\bibfnamefont{E.}~\bibnamefont{Peik}}, \bibnamefont{and}
  \bibinfo{author}{\bibfnamefont{P.~O.} \bibnamefont{Schmidt}},
  \bibinfo{journal}{Reviews of Modern Physics} \textbf{\bibinfo{volume}{87}},
  \bibinfo{pages}{637} (\bibinfo{year}{2015}).

\bibitem[{\citenamefont{Tr{\"a}bert et~al.}(2003)\citenamefont{Tr{\"a}bert,
  Calamai, Gwinner, Knystautas, Pinnington, and Wolf}}]{trabert2003m1}
\bibinfo{author}{\bibfnamefont{E.}~\bibnamefont{Tr{\"a}bert}},
  \bibinfo{author}{\bibfnamefont{A.}~\bibnamefont{Calamai}},
  \bibinfo{author}{\bibfnamefont{G.}~\bibnamefont{Gwinner}},
  \bibinfo{author}{\bibfnamefont{E.}~\bibnamefont{Knystautas}},
  \bibinfo{author}{\bibfnamefont{E.}~\bibnamefont{Pinnington}},
  \bibnamefont{and} \bibinfo{author}{\bibfnamefont{A.}~\bibnamefont{Wolf}},
  \bibinfo{journal}{Journal of Physics B: Atomic, Molecular and Optical
  Physics} \textbf{\bibinfo{volume}{36}}, \bibinfo{pages}{1129}
  (\bibinfo{year}{2003}).

\bibitem[{\citenamefont{Beiersdorfer}(2009)}]{beiersdorfer2009spectroscopy}
\bibinfo{author}{\bibfnamefont{P.}~\bibnamefont{Beiersdorfer}},
  \bibinfo{journal}{Physica Scripta} \textbf{\bibinfo{volume}{2009}},
  \bibinfo{pages}{014010} (\bibinfo{year}{2009}).

\bibitem[{\citenamefont{Windberger et~al.}(2015)\citenamefont{Windberger,
  L{\'o}pez-Urrutia, Bekker, Oreshkina, Berengut, Bock, Borschevsky, Dzuba,
  Eliav, Harman et~al.}}]{windberger2015identification}
\bibinfo{author}{\bibfnamefont{A.}~\bibnamefont{Windberger}},
  \bibinfo{author}{\bibfnamefont{J.~C.} \bibnamefont{L{\'o}pez-Urrutia}},
  \bibinfo{author}{\bibfnamefont{H.}~\bibnamefont{Bekker}},
  \bibinfo{author}{\bibfnamefont{N.}~\bibnamefont{Oreshkina}},
  \bibinfo{author}{\bibfnamefont{J.}~\bibnamefont{Berengut}},
  \bibinfo{author}{\bibfnamefont{V.}~\bibnamefont{Bock}},
  \bibinfo{author}{\bibfnamefont{A.}~\bibnamefont{Borschevsky}},
  \bibinfo{author}{\bibfnamefont{V.}~\bibnamefont{Dzuba}},
  \bibinfo{author}{\bibfnamefont{E.}~\bibnamefont{Eliav}},
  \bibinfo{author}{\bibfnamefont{Z.}~\bibnamefont{Harman}},
  \bibnamefont{et~al.}, \bibinfo{journal}{Physical Review Letters}
  \textbf{\bibinfo{volume}{114}}, \bibinfo{pages}{150801}
  (\bibinfo{year}{2015}).

\bibitem[{\citenamefont{Safronova et~al.}(2017)\citenamefont{Safronova,
  Safronova, and Johnson}}]{safronova2017forbidden}
\bibinfo{author}{\bibfnamefont{U.}~\bibnamefont{Safronova}},
  \bibinfo{author}{\bibfnamefont{M.}~\bibnamefont{Safronova}},
  \bibnamefont{and} \bibinfo{author}{\bibfnamefont{W.}~\bibnamefont{Johnson}},
  \bibinfo{journal}{Physical Review A} \textbf{\bibinfo{volume}{95}},
  \bibinfo{pages}{042507} (\bibinfo{year}{2017}).

\bibitem[{\citenamefont{Mrozowski}(1944)}]{mrozowski1944forbidden}
\bibinfo{author}{\bibfnamefont{S.}~\bibnamefont{Mrozowski}},
  \bibinfo{journal}{Reviews of Modern Physics} \textbf{\bibinfo{volume}{16}},
  \bibinfo{pages}{153} (\bibinfo{year}{1944}).

\bibitem[{\citenamefont{Kwela et~al.}(1982)\citenamefont{Kwela, Kowalski, and
  Heldt}}]{kwela1982determination}
\bibinfo{author}{\bibfnamefont{J.}~\bibnamefont{Kwela}},
  \bibinfo{author}{\bibfnamefont{A.}~\bibnamefont{Kowalski}}, \bibnamefont{and}
  \bibinfo{author}{\bibfnamefont{J.}~\bibnamefont{Heldt}},
  \bibinfo{journal}{Journal of Optical Society of America}
  \textbf{\bibinfo{volume}{72}}, \bibinfo{pages}{1550} (\bibinfo{year}{1982}).

\bibitem[{\citenamefont{Augustyniak et~al.}(1975)\citenamefont{Augustyniak,
  Heldt, and Bronowski}}]{augustyniak1975zeeman}
\bibinfo{author}{\bibfnamefont{L.}~\bibnamefont{Augustyniak}},
  \bibinfo{author}{\bibfnamefont{J.}~\bibnamefont{Heldt}}, \bibnamefont{and}
  \bibinfo{author}{\bibfnamefont{J.}~\bibnamefont{Bronowski}},
  \bibinfo{journal}{Physica Scripta} \textbf{\bibinfo{volume}{12}},
  \bibinfo{pages}{157} (\bibinfo{year}{1975}).

\bibitem[{\citenamefont{Werbowy and Kwela}(2009)}]{werbowy2009m1}
\bibinfo{author}{\bibfnamefont{S.}~\bibnamefont{Werbowy}} \bibnamefont{and}
  \bibinfo{author}{\bibfnamefont{J.}~\bibnamefont{Kwela}},
  \bibinfo{journal}{Journal of Physics B: Atomic, Molecular and Optical
  Physics} \textbf{\bibinfo{volume}{42}}, \bibinfo{pages}{065002}
  (\bibinfo{year}{2009}).

\bibitem[{\citenamefont{Wa{\c{s}}owicz}(2007)}]{wacsowicz2007e2}
\bibinfo{author}{\bibfnamefont{T.}~\bibnamefont{Wa{\c{s}}owicz}},
  \bibinfo{journal}{Physica Scripta} \textbf{\bibinfo{volume}{76}},
  \bibinfo{pages}{294} (\bibinfo{year}{2007}).

\bibitem[{\citenamefont{Van~Enk and Nienhuis}(1994)}]{van1994spin}
\bibinfo{author}{\bibfnamefont{S.}~\bibnamefont{Van~Enk}} \bibnamefont{and}
  \bibinfo{author}{\bibfnamefont{G.}~\bibnamefont{Nienhuis}},
  \bibinfo{journal}{European Physics Letters} \textbf{\bibinfo{volume}{25}},
  \bibinfo{pages}{497} (\bibinfo{year}{1994}).

\bibitem[{\citenamefont{Durnin et~al.}(1987)\citenamefont{Durnin, Miceli~Jr,
  and Eberly}}]{durnin1987diffraction}
\bibinfo{author}{\bibfnamefont{J.}~\bibnamefont{Durnin}},
  \bibinfo{author}{\bibfnamefont{J.}~\bibnamefont{Miceli~Jr}},
  \bibnamefont{and} \bibinfo{author}{\bibfnamefont{J.}~\bibnamefont{Eberly}},
  \bibinfo{journal}{Physical Review Letters} \textbf{\bibinfo{volume}{58}},
  \bibinfo{pages}{1499} (\bibinfo{year}{1987}).

\bibitem[{\citenamefont{Born and Wolf}(2013)}]{born2013principles}
\bibinfo{author}{\bibfnamefont{M.}~\bibnamefont{Born}} \bibnamefont{and}
  \bibinfo{author}{\bibfnamefont{E.}~\bibnamefont{Wolf}},
  \emph{\bibinfo{title}{Principles of optics: electromagnetic theory of
  propagation, interference and diffraction of light}}
  (\bibinfo{publisher}{Elsevier}, \bibinfo{year}{2013}).

\bibitem[{\citenamefont{Jentschura and Serbo}(2011)}]{jentschura2011generation}
\bibinfo{author}{\bibfnamefont{U.}~\bibnamefont{Jentschura}} \bibnamefont{and}
  \bibinfo{author}{\bibfnamefont{V.}~\bibnamefont{Serbo}},
  \bibinfo{journal}{Physical Review Letters} \textbf{\bibinfo{volume}{106}},
  \bibinfo{pages}{013001} (\bibinfo{year}{2011}).

\bibitem[{\citenamefont{Bliokh et~al.}(2010)\citenamefont{Bliokh, Alonso,
  Ostrovskaya, and Aiello}}]{bliokh2010angular}
\bibinfo{author}{\bibfnamefont{K.~Y.} \bibnamefont{Bliokh}},
  \bibinfo{author}{\bibfnamefont{M.~A.} \bibnamefont{Alonso}},
  \bibinfo{author}{\bibfnamefont{E.~A.} \bibnamefont{Ostrovskaya}},
  \bibnamefont{and} \bibinfo{author}{\bibfnamefont{A.}~\bibnamefont{Aiello}},
  \bibinfo{journal}{Physical Review A} \textbf{\bibinfo{volume}{82}},
  \bibinfo{pages}{063825} (\bibinfo{year}{2010}).

\bibitem[{\citenamefont{Bliokh et~al.}(2011)\citenamefont{Bliokh, Ostrovskaya,
  Alonso, Rodr{\'\i}guez-Herrera, Lara, and Dainty}}]{bliokh2011spin}
\bibinfo{author}{\bibfnamefont{K.~Y.} \bibnamefont{Bliokh}},
  \bibinfo{author}{\bibfnamefont{E.~A.} \bibnamefont{Ostrovskaya}},
  \bibinfo{author}{\bibfnamefont{M.~A.} \bibnamefont{Alonso}},
  \bibinfo{author}{\bibfnamefont{O.~G.} \bibnamefont{Rodr{\'\i}guez-Herrera}},
  \bibinfo{author}{\bibfnamefont{D.}~\bibnamefont{Lara}}, \bibnamefont{and}
  \bibinfo{author}{\bibfnamefont{C.}~\bibnamefont{Dainty}},
  \bibinfo{journal}{Optics Express} \textbf{\bibinfo{volume}{19}},
  \bibinfo{pages}{26132} (\bibinfo{year}{2011}).

\bibitem[{\citenamefont{Kaplan and McGuire}(2015)}]{Kaplan15}
\bibinfo{author}{\bibfnamefont{L.}~\bibnamefont{Kaplan}} \bibnamefont{and}
  \bibinfo{author}{\bibfnamefont{J.~H.} \bibnamefont{McGuire}},
  \bibinfo{journal}{Phys. Rev. A} \textbf{\bibinfo{volume}{92}},
  \bibinfo{pages}{032702} (\bibinfo{year}{2015}),
  \urlprefix\url{http://link.aps.org/doi/10.1103/PhysRevA.92.032702}.

\bibitem[{\citenamefont{Sheppard and Wilson}(1978)}]{sheppard1978gaussian}
\bibinfo{author}{\bibfnamefont{C.}~\bibnamefont{Sheppard}} \bibnamefont{and}
  \bibinfo{author}{\bibfnamefont{T.}~\bibnamefont{Wilson}},
  \bibinfo{journal}{IEE Journal on Microwaves, Optics and Acoustics}
  \textbf{\bibinfo{volume}{2}}, \bibinfo{pages}{105} (\bibinfo{year}{1978}).

\bibitem[{\citenamefont{Kiselev}(2004)}]{kiselev2004new}
\bibinfo{author}{\bibfnamefont{A.}~\bibnamefont{Kiselev}},
  \bibinfo{journal}{Optics and Spectroscopy} \textbf{\bibinfo{volume}{96}},
  \bibinfo{pages}{479} (\bibinfo{year}{2004}).

\bibitem[{\citenamefont{Andrews and Babiker}(2012)}]{andrews2012angular}
\bibinfo{author}{\bibfnamefont{D.~L.} \bibnamefont{Andrews}} \bibnamefont{and}
  \bibinfo{author}{\bibfnamefont{M.}~\bibnamefont{Babiker}},
  \emph{\bibinfo{title}{The angular momentum of light}}
  (\bibinfo{publisher}{Cambridge University Press}, \bibinfo{year}{2012}).

\bibitem[{\citenamefont{Akhiezer and Berestetskii}(1959)}]{akhiezer1959quantum}
\bibinfo{author}{\bibfnamefont{A.}~\bibnamefont{Akhiezer}} \bibnamefont{and}
  \bibinfo{author}{\bibfnamefont{V.}~\bibnamefont{Berestetskii}},
  \emph{\bibinfo{title}{Quantum Electrodynamics, 1965}}
  (\bibinfo{publisher}{Interscience, New York}, \bibinfo{year}{1959}).

\bibitem[{\citenamefont{Afanasev et~al.}(2017)\citenamefont{Afanasev, Carlson,
  and Solyanik}}]{afanasev2017circular}
\bibinfo{author}{\bibfnamefont{A.}~\bibnamefont{Afanasev}},
  \bibinfo{author}{\bibfnamefont{C.~E.} \bibnamefont{Carlson}},
  \bibnamefont{and} \bibinfo{author}{\bibfnamefont{M.}~\bibnamefont{Solyanik}},
  \bibinfo{journal}{Journal of Optics} \textbf{\bibinfo{volume}{19}},
  \bibinfo{pages}{105401} (\bibinfo{year}{2017}),
  \urlprefix\url{http://stacks.iop.org/2040-8986/19/i=10/a=105401}.

\bibitem[{\citenamefont{Rynkun et~al.}(2012)\citenamefont{Rynkun, J{\"o}nsson,
  Gaigalas, and Fischer}}]{rynkun2012energies}
\bibinfo{author}{\bibfnamefont{P.}~\bibnamefont{Rynkun}},
  \bibinfo{author}{\bibfnamefont{P.}~\bibnamefont{J{\"o}nsson}},
  \bibinfo{author}{\bibfnamefont{G.}~\bibnamefont{Gaigalas}}, \bibnamefont{and}
  \bibinfo{author}{\bibfnamefont{C.~F.} \bibnamefont{Fischer}},
  \bibinfo{journal}{Atomic Data and Nuclear Data Tables}
  \textbf{\bibinfo{volume}{98}}, \bibinfo{pages}{481} (\bibinfo{year}{2012}).

\bibitem[{\citenamefont{Bahrdt et~al.}(2013)\citenamefont{Bahrdt, Holldack,
  Kuske, M\"uller, Scheer, and Schmid}}]{BESSY99ev}
\bibinfo{author}{\bibfnamefont{J.}~\bibnamefont{Bahrdt}},
  \bibinfo{author}{\bibfnamefont{K.}~\bibnamefont{Holldack}},
  \bibinfo{author}{\bibfnamefont{P.}~\bibnamefont{Kuske}},
  \bibinfo{author}{\bibfnamefont{R.}~\bibnamefont{M\"uller}},
  \bibinfo{author}{\bibfnamefont{M.}~\bibnamefont{Scheer}}, \bibnamefont{and}
  \bibinfo{author}{\bibfnamefont{P.}~\bibnamefont{Schmid}},
  \bibinfo{journal}{Phys. Rev. Lett.} \textbf{\bibinfo{volume}{111}},
  \bibinfo{pages}{034801} (\bibinfo{year}{2013}),
  \urlprefix\url{https://link.aps.org/doi/10.1103/PhysRevLett.111.034801}.

\end{thebibliography}

\end{document}